\documentclass[twocolumn, reprint,showpacs, superscriptaddress, amsmath,amssymb, aps, prl]{revtex4-1}

\usepackage{graphicx}
\usepackage{amsmath}
\usepackage{epsfig}
\usepackage{color}
\usepackage[english]{babel}

\usepackage{dcolumn}% Align table columns on decimal point
\usepackage{bm}% bold math
\usepackage[
   bookmarks=true,
   colorlinks=true,
   hyperindex=true,
   citecolor=blue,
   linkcolor=blue,
   urlcolor=blue
]{hyperref}

\begin{document}

\title{Coherence preservation of a single neutral atom qubit transferred between magic-intensity optical traps }

\author{Jiaheng Yang}
\affiliation{State Key Laboratory of Magnetic Resonance and Atomic and Molecular Physics, Wuhan
Institute of Physics and Mathematics, Chinese Academy of Sciences - Wuhan National
Laboratory for Optoelectronics, Wuhan 430071, China}
\affiliation{School of Physics, University of Chinese Academy of Sciences, Beijing 100049, China}

\author{Xiaodong He}
\email{hexd@wipm.ac.cn}
\affiliation{State Key Laboratory of Magnetic Resonance and Atomic and Molecular Physics, Wuhan
Institute of Physics and Mathematics, Chinese Academy of Sciences - Wuhan National
Laboratory for Optoelectronics, Wuhan 430071, China}
\affiliation{Center for Cold Atom Physics, Chinese Academy of Sciences, Wuhan 430071, China}

\author{Ruijun Guo}
\affiliation{State Key Laboratory of Magnetic Resonance and Atomic and Molecular Physics, Wuhan
Institute of Physics and Mathematics, Chinese Academy of Sciences - Wuhan National
Laboratory for Optoelectronics, Wuhan 430071, China}
\affiliation{School of Physics, University of Chinese Academy of Sciences, Beijing 100049, China}

\author{Peng Xu}
%\email{etherxp@wpim.ac.cn}
\affiliation{State Key Laboratory of Magnetic Resonance and Atomic and Molecular Physics, Wuhan
Institute of Physics and Mathematics, Chinese Academy of Sciences - Wuhan National
Laboratory for Optoelectronics, Wuhan 430071, China}
\affiliation{Center for Cold Atom Physics, Chinese Academy of Sciences, Wuhan 430071, China}

\author{Kunpeng Wang}
\affiliation{State Key Laboratory of Magnetic Resonance and Atomic and Molecular Physics, Wuhan
Institute of Physics and Mathematics, Chinese Academy of Sciences - Wuhan National
Laboratory for Optoelectronics, Wuhan 430071, China}
\affiliation{School of Physics, University of Chinese Academy of Sciences, Beijing 100049, China}

\author{Cheng Sheng}
\affiliation{State Key Laboratory of Magnetic Resonance and Atomic and Molecular Physics, Wuhan
Institute of Physics and Mathematics, Chinese Academy of Sciences - Wuhan National
Laboratory for Optoelectronics, Wuhan 430071, China}

\author{Min Liu}
\affiliation{State Key Laboratory of Magnetic Resonance and Atomic and Molecular Physics, Wuhan
Institute of Physics and Mathematics, Chinese Academy of Sciences - Wuhan National
Laboratory for Optoelectronics, Wuhan 430071, China}
\affiliation{Center for Cold Atom Physics, Chinese Academy of Sciences, Wuhan 430071, China}

\author{Jin Wang}
\affiliation{State Key Laboratory of Magnetic Resonance and Atomic and Molecular Physics, Wuhan
Institute of Physics and Mathematics, Chinese Academy of Sciences - Wuhan National
Laboratory for Optoelectronics, Wuhan 430071, China}
\affiliation{Center for Cold Atom Physics, Chinese Academy of Sciences, Wuhan 430071, China}

\author{Andrei Derevianko}
\affiliation{Department of Physics, University of Nevada, Reno, Nevada 89557, USA}
\affiliation{Center for Cold Atom Physics, Chinese Academy of Sciences, Wuhan 430071, China}

\author{Mingsheng Zhan}
\email{mszhan@wipm.ac.cn}
\affiliation{State Key Laboratory of Magnetic Resonance and Atomic and Molecular Physics, Wuhan
Institute of Physics and Mathematics, Chinese Academy of Sciences - Wuhan National
Laboratory for Optoelectronics, Wuhan 430071, China}
\affiliation{Center for Cold Atom Physics, Chinese Academy of Sciences, Wuhan 430071, China}

\date{\today}

\begin{abstract}
We demonstrate that the coherence of a single mobile atomic qubit can be well preserved during a transfer process among different optical dipole traps (ODTs). This is a prerequisite step in realizing a large-scale neutral atom quantum information processing platform.
A qubit encoded in the hyperfine manifold  of $^{87}$Rb atom  is dynamically extracted from the static quantum register by an auxiliary moving ODT and reinserted into the static ODT.  Previous experiments were limited by decoherences induced by the differential light shifts of qubit states. Here we apply a magic-intensity trapping technique which mitigates the detrimental effects of light shifts and substantially enhances the coherence time to $225 \pm 21\,\mathrm{ms}$. The experimentally demonstrated magic trapping technique relies on the previously neglected hyperpolarizability contribution to the light shifts, which makes the light shift dependence on the trapping laser intensity to be parabolic. Because of the parabolic dependence, at a certain ``magic'' intensity, the first order sensitivity to trapping light intensity variations over ODT volume is eliminated.  We experimentally demonstrate the utility of this approach and measure  hyperpolarizability for the first time. Our results pave the way for constructing a scalable quantum-computing architectures with single atoms trapped in an array of magic ODTs.
\end{abstract}


%37.10.Jk	Atoms in optical lattices
%03.67.Lx	Quantum computation architectures and implementations
%42.50.Ct	Quantum description of interaction of light and matter; related experiments

\maketitle

%%%%% [Motivations]
%%%%%%%%%%%%%%%%%%%%%%%%%%%%%%%%%%%%%%%%%

A quantum computer~\cite{DiVincenzo2000} or a simulator is a scalable physical system with coherently controllable and well characterized qubits.  As an important candidate for quantum information processing and quantum simulation, a microscopic array of  single atoms confined in optical dipole traps (ODTs) has attracted a great deal of interest in recent years~\cite{Saffman2010,Georgescu2014}.
In such architectures~\cite{Weitenberg2011} each ODT-stored atom acts as a qubit, and an array of single atoms in static ODTs forms a quantum register. An important requirement is the ability  to controllably transport a remote qubit, acting as a mobile qubit, into the interaction range with other register atoms  for performing two-qubit gates. This transfer must be carried out without influencing other qubits of the large-scale quantum register. Recently, we experimentally demonstrated such a transfer scheme~\cite{Yu2014}, in which the single mobile qubit was dynamically extracted from a ring optical lattice site by an auxiliary moving ODT and reinserted into the original site.
We, however, found that during the transfer process the qubits severely lose coherence.
Although an alternative transfer scheme between two ODTs has been also demonstrated ~\cite{Beugnon2007} and the coherence of the mobile qubit was found not to be affected during the transfer, this scheme is not suitable for scalable quantum systems because the register static ODTs are switched off during the transfer. If the register keeps holding qubits as required for a scalable system, the static ODTs should remain always on. Then the mobile qubit unavoidably experiences large variations of the trapping potential in the merging process between moving and static ODTs, leading to the coherence losses.

Typically, an atomic qubit is encoded into a superposition of two hyperfine Zeeman levels of the ground states of an alkali-metal atom. Generically different hyperfine states experience mismatched light shifts induced by the trapping laser field, leading to the so-called differential light shift (DLS). The DLS depends on the laser intensity at the qubit position and due to the spatial distribution of laser field intensity in a trap,  the qubit suffers from strong inhomogeneous dephasing effect. Thereby the coherence time is limited to scales of several ms in red-detuned ODTs~\cite{Kuhr2005,Yavuz2006,Jones2007,Yu2013}, or several tens of ms in blue-detuned ODTs~\cite{Xu2010,Li2012}. To reduce the DLS-induced dephasing, one could add a weak near-resonant compensating laser beam, but at an expense of a substantially increased scattering rate~\cite{Kaplan2002,Chaudhury2006}, or employ the dynamical decoupling methods such as the spin echo or the Carr-Purcell-Meiboom-Gill sequence~\cite{Kuhr2005,Yu2013}. The dynamical decoupling methods are found to be efficient for qubits in static ODTs but inefficient for mobile qubits.  Indeed, the heating of atoms and pointing instabilities of the trap laser beams during the transfer can not be efficiently suppressed by the dynamical decoupling methods, causing the mobile qubits to lose coherence.
 %{\bf do you need this? even though the whole transfer process is implemented carefully~\cite{Yu2014}}.

Similar to optical lattice clocks~\cite{Katori2003}, a complete control approach over DLS is to construct a ``magic'' trap, where the two qubit states experience identical trapping potentials and the relative phase accumulation is nearly independent of the atomic center-of-mass motion and  trapping field fluctuations. To this end, exploiting the vector light shift, which acts like an effective Zeeman field $B_\mathrm{eff}$, to zero out the DLS of $m_F\neq0$ hyperfine states has been proposed~\cite{Jai2007,Flambaum2008} and demonstrated in $^{7}$Li~\cite{Kim2013}. Similarly exploiting the vector light shift for cancelling DLS of $m_F$=0 hyperfine states in $^{87}$Rb atoms has also been demonstrated~\cite{Lundblad2010,Derevianko2010}. While at the cost of increased sensitivity to the magnetic noise due to the requirement of a several Gauss magnetic bias field, this technique has been proven to be efficient in enhancing the lifetime of spin-wave qubits in a $^{87}$Rb ensemble~\cite{Dudin2010PRA,Dudin2010PRL}. Furthermore, to reduce the sensitivity to fluctuations of both laser and magnetic fields, doubly magic trapping for $m_F\neq0$ state was proposed~\cite{Derevianko2010PRL} and experimentally demonstrated in $^{87}$Rb atoms confined in optical lattice~\cite{Chicireanu2011}. To date, the magic trapping techniques have been proved to be efficient in suppressing inhomogeneous DLS of atoms in static ODTs. The open question is whether these technique can be also used to mitigate coherence loss in manipulating the mobile qubits. This question is explicitly answered in this Letter.

We begin by studying the DLS of solitaty stationary $^{87}$Rb qubits (here $|0\rangle\equiv|F=1,m_F=0\rangle$ and $|1\rangle\equiv|F=2,m_F=0\rangle$) confined in a circularly polarized ODT. We observe and measure previously neglected ground state hyperpolarizability, which makes the DLS dependence on laser intensity to be parabolic. Because of the parabolic dependence, at a certain ``magic'' intensity,  the first order sensitivity to trapping light intensity variations is eliminated~\cite{Carr2014}. We further demonstrate that the measured coherence time of the mobile qubits is the same as for the static qubits, i.e., the transfer process does not induce extra coherence loss.

%%%%%%%%%%%%%%%%%%%%%%%%%%%
\begin{figure}[htbp]
\centering
\includegraphics[width=8.6cm]{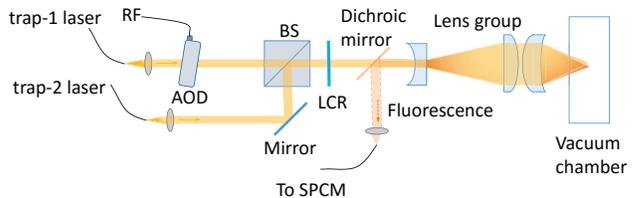}
\caption{(Color online) Schematic of the optical layout. A movable 830 nm light beam (labeled as trap-1) is deflected in two orthogonal directions by an acoustic-optic deflector (AOD) which is driven by a radio-frequency (RF) signal. The trap-1 is combined with another 830 nm light beam (trap-2) by a beam splitter (BS). Their polarizations are purified by a Glan-Thompson polarizer first, then actively controlled by a liquid crystal retarder (LCR). Both laser beams are finally focused by a microscopic objective to provide a 3D confinement. The same objective also collects fluorescence from the trapped atoms. The fluorescence is then detected by a single photon counting module (SPCM).}
\label{fig:fig1}
\end{figure}
%%%%%%%%%%%%%%%%%%%%%%%%%%%

The experimental details on trapping single $^{87}$Rb atoms and individual qubit manipulations have been descried elsewhere~\cite{Yu2013,Yu2014}. Here, a modified optical layout is  illustrated in Fig.~\ref{fig:fig1}. The  wavelength of the dipole laser beams is $\lambda=830.0$ nm with the waist of the trap-2 of 1.25 $\mu$m.
We load a single $^{87}$Rb atom from a magneto-optical trap via a collisional blockade mechanism~\cite{Schlosser2002}. It is worth noting that in previous experiments on manipulating degenerate ensembles in optical lattice~\cite{Lundblad2010}, the trap depth $U_{a}\approx3.5\ \mu$K and thereby $B_\mathrm{eff} = 12\,\mathrm{mG}$ can be neglected to the bias B-field. But here we confine single atoms with temperature of several tens of $\mu$K in an ODT with a much larger trap depth up to $0.6$ mK. Now the $B_\mathrm{eff}\approx$ 1.120 G becomes comparable to the externally applied B-field. The corresponding vector light shift is so strong that the usually neglected  ground state hyperpolarizability becomes important and must  be taken into account. Recent  theoretical analysis by Carr and Saffman~\cite{Carr2014} revealed  the importance of hyperpolarizability in reaching magic conditions in trapping of Cs atoms.

The DLS of Zeeman-insensitive clock transition experienced  by the $^{87}$Rb atoms in a magnetic field $B$ reads
\begin{equation}
 \label{eq1}
 \delta\nu(B,U_a)=\beta_1 U_a+\beta_2 B U_a+\beta_4 U_a^2,
\end{equation}
where $\delta\nu$ is the total DLS seen by the atoms, $U_a$ (in unit of Hz) is the local trap depth,  $\beta_1$ is the coefficient of the third order hyperfine-mediated polarizability, $\beta_2$ is the coefficient of the third order cross-term and $\beta_4$ is the coefficient of the ground state hyperpolarizability. The local trap depth $U_{a}$ is given in terms of the dominant dynamic ground-state polarizability $\alpha_{5s}\left(\omega\right)$  and the local E-field amplitude $\mathcal{E}_{a}$ as
$U_{a}=-\alpha_{5s}\left(\omega\right)\left(\mathcal{E}_{a}/2\right)^{2}/h$,
where $h$ is the Planck constant. Then $\beta_{1}=\left[
\alpha_{F=2}\left(  \omega\right)  -\alpha_{F=1}\left(  \omega\right)
\right]  /\alpha_{5s}\left(  \omega\right)  $, where $\alpha_{F}\left(
\omega\right)  $ are the hyperfine--interaction-mediated
polarizabilities~\cite{Derevianko2010,Carr2014} of hyperfine sublevels.
$\beta_{2}=-2 A\mu_{B}\alpha_{5s}^{a}(\omega)[h\nu_{0}\alpha_{5s}\left(
\omega\right)]^{-1}$
is expressed in terms of the degree of circular polarization
$A$, hyperfine splitting $\nu_{0}$ and the vector polarizability of the ground
state $\alpha_{5s}^{a}\left(  \omega\right)$.
Finally,
 $\beta_{4}=(A^{2}/2\nu_{0})[\alpha_{5s}^{a}\left(  \omega\right)
/\alpha_{5s}\left(  \omega\right)] ^{2}$.

%%%%%%%%%%%%%%%%%%%%%%%%%%%
\begin{figure}[htbp]
\centering
\includegraphics[width=8.0cm]{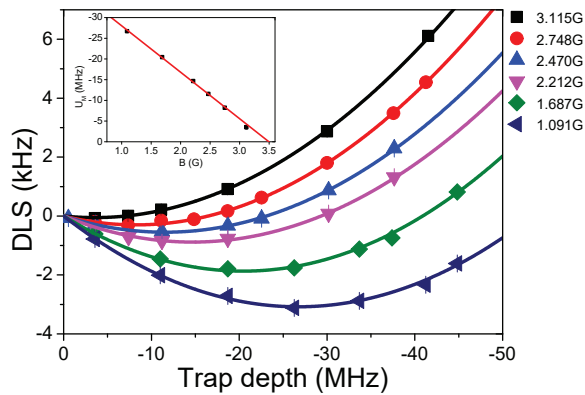}
\caption{(Color online) DLS in the presence of hyperpolarizability. DLS of a qubit in the circularly polarized trap-2 is measured as a function of trap depthes at various magnetic field strengthes. The solid curves are fits to the Eq.(1). The inset plots the minima $U_M$ in the DLS curves as a function of magnetic field $B$. The light intensity of each minimum is chosen  as the magic intensity at that B-field value.}
\label{fig:fig2}
\end{figure}
%%%%%%%%%%%%%%%%%%%%%%%%%%%%%%%%%%%%%%%%%%%%%%%%%%%%%%%%%%%

$\beta_2$ and $\beta_4$ depend on the degree of the circular polarization. For the sake of simplicity, we use fully circular $\sigma^{+}$ light, that is $A=1$. Varying the trap depths and B-fields  we can deduce the values of $\beta_2$ and $\beta_4$  in Eq.~(\ref{eq1}) from our DLS measurements. In case of linearly polarized light field,  $\beta_2$ and $\beta_4$ terms vanish and DLS is linearly dependent on the trap depth. Thereby we calibrate the trap depth by comparing the measured DLS in the linearly polarized trap with the calculated value of $\beta_{1} \approx 3.67 \times10^{-4}$ from the atomic structure data~\cite{Fam2013,Shih2013}. Then we measure the DLS curves in the circularly polarized trap-2 for several values of magnetic fields. As shown in Fig.~\ref{fig:fig2}, all of the measured curves exhibit nonlinear (parabolic) dependence of the DLS on the trap depths unlike the linear dependence in previous measurements~\cite{Lundblad2010}. Given our calculated value of $\beta_{1}\approx3.47\times10^{-4}$ for circular polarization, all the curves are fitted to Eq.(\ref{eq1}) yielding the values of $\beta_2$ and $\beta_4$.  Averaging over all of the fitted results, the $\beta_2$ and $\beta_4$ are found to be -0.99(3)$\times10^{-4}\, \mathrm{G}^{-1}$ and 4.6(2)$\times$10$^{-12} \,\mathrm{Hz}^{-1}$ respectively. We also carried out numerical evaluation of $\beta$ parameters using the formalism and the high-accuracy techniques of atomic structure described in Refs.~\cite{Derevianko2010, Derevianko2010PRL}. The theoretical results, $\beta_{2}=-1.03\times10^{-4}~\mathrm{G}^{-1}$ and $\beta_{4}=4.64\times10^{-12}~\mathrm{Hz}^{-1}$, are in a good agreement with the experimental values. Further, from Eq.~(1), the minimum trap depths are given by $U_M={-(\beta_1+\beta_2 B)}{(2\beta_4)^{-1}}$, i.e., they scale linearly with B-field.
Fig.~\ref{fig:fig2} inset shows the linear dependence of the measured DLS minima on the external B-field. When $B \rightarrow-{\beta_1}/{\beta_2}\approx 3.51\, \mathrm{G}$,  $U_M$ approaches 0 and the trap is too weak to trap atoms. In contrast, for smaller B-fields, larger magic light intensity trapping depths are needed. It means that the atoms scatter more spontaneous Raman photons from the trapping laser, leading to faster spin relaxation rate. To strike a compromise between the reliably of trapping atoms and the suppression of the spin relaxation rate, we set the working magnetic field  to  $3.115 \,\mathrm{G}$.

\begin{figure}[htbp]
\centering
\includegraphics[width=8.0cm]{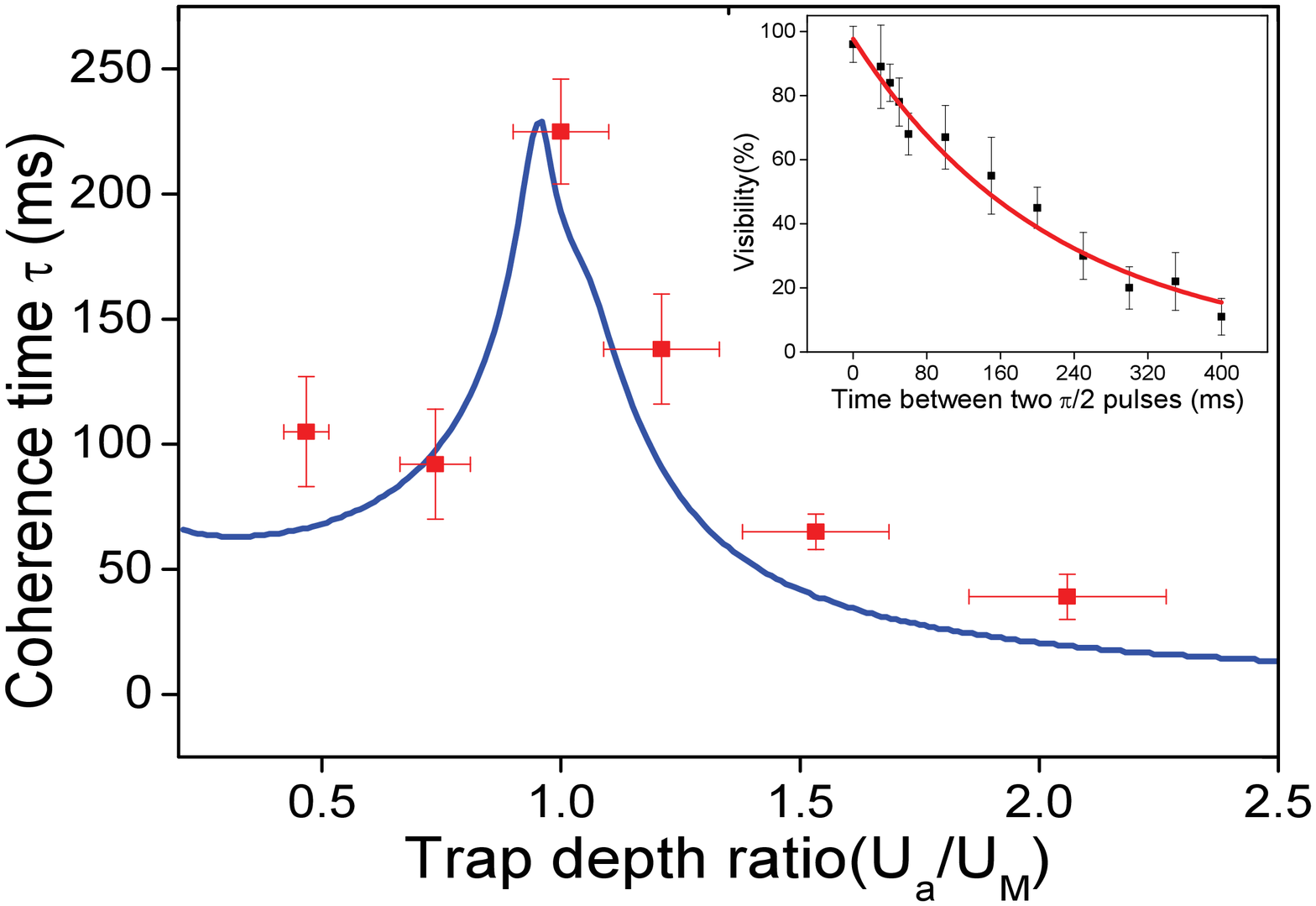}
\caption{(Color online) Coherence time $\tau$ and its dependence on normalized ratios $U_a/U_M$. At $U_a = U_M, \tau = 225 \pm 21 \,\mathrm{ms}$. The error bars of ratios are from the measured error of $U_M$ (10$\%$). A coherence time is extracted from a decay time of the envelope of Ramsey visibility, as shown in the inset, which is the measured visibility of Ramsey signals as a function of the duration between two $\pi/2$ pulses at $U_M$. All the accompanying error bars of coherence times and visibility are fitting errors. The theoretical curve is obtained by combining the calculated $T_2^*$ with Eq.(2), an estimated $T_2'\approx300$ ms and the measured value of $T_1\approx4$ s. }
\label{fig:visi}
\end{figure}

Next we measure the dependence of qubit coherence times on the ratios of trap depth to the measured magic trap depth which is the fitted minimum (with $10\%$ uncertainty) in the DLS curves for $3.115 \,\mathrm{G}$ in trap-2. The coherence time is measured by recording the decay of the visibility of Ramsey signal, as shown in the inset of Fig.~\ref{fig:visi}. By varying the trap depths, we find the longest  coherence time at around $U_M$, which is consistent with the magic operating condition $\partial\delta\nu(B_0,U_a) / \partial U_{a}=0$. At $U_a = U_M, \tau = 225 \pm 21 \, \mathrm{ms}$.

%%%%%%%%%%%%%%%%%%%%%%%%%%%
%%%%%%%%%%%%%%%%%%%%%%%%%%%

Theoretical interpretation of Ramsey-type measurements (solid curve in Fig.~\ref{fig:visi}) requires special care due to the thermal distribution of loaded atoms. Before proceeding with the thermal averaging, we remind the reader that the
decay time $\tau$ of the Ramsey signal can be decomposed into two main parts, ${1}/{\tau}={1}/{T_1}+{1}/{T_2}$, where $T_1$ is longitudinal relaxation time and $T_2$ is transverse decay time. In our experiment, the measured $T_1$ is over 4 s and ${1}/{T_1}$ can be neglected. In addition, $T_2$ can be  decomposed as ${1}/{T_2}={1}/{T_2^\prime}+{1}/{T_2^*}$, where $T_2^\prime$ is the homogeneous dephasing time and $T_2^*$ is the inhomogeneous reversible dephasing time~\cite{Kuhr2005,Yu2013}.

 In our experiments, the DLS and coherence time are deduced from multiple repeated single atom measurements. We thereby consider a collection of all trapped single atoms as a thermal ensemble. Their energy distribution in the ODT is the Boltzmann distribution with the probability density
$p(E_a)=(E_a/\sqrt{2})^2(k_B T_a)^{-3}\exp(-E_a/k_B T_a)~$\cite{Kuhr2005},
where $k_B$ is the Boltzmann constant, $E_a$ is the sum of kinetic and potential energies, and $T_a$ is the temperature of the atoms.
As a thermal ensemble, atoms have an average potential energy $3k_B T_a/2$.
The average trap depth $U_a$ seen by the thermal atoms fulfills $hU_0= hU_a-3k_B T_a/2$, where $U_0$ is the trap depth at trap minimum.
%{\bf the trap potential located at the bottom of the trap??}
%{\bf $\leftarrow$ what did you mean here? rewrite}.
Thereby the average trap depth $U(E_a)$ of an atom of energy $E_a>0$
is given by $U(E_a)=U_0+E_a/(2h)$ in the harmonic approximation.
For magic trapping $U_a(E_a)=U_M$ and with Eq.~(1) it reduces to a DLS $\delta\nu(B,U_M)=\delta\nu_M+\delta\nu_E$,
where $\delta\nu_M=-{(\beta_1+B\beta_2)^2}{(4\beta_4)^{-1}}$ is the DLS minimum, and $\delta\nu_E={\beta_4(E_a-3 k_B T_a)^2}{(2h)^{-2}}$. Clearly the temperature-dependent term $\delta\nu_E$ leads to residual inhomogeneous reversible dephasing.

Further, the Ramsey interrogation sequence consists of applying two $\pi/2$ microwave pulses separated by a time interval $t$. The population of the $|0\rangle$ qubit state at the end of the sequence follows
 $P_0(t)=1/2+\cos[2\pi(\delta+\delta\nu(B,U))t]/2$, where $\delta$ is the detuning of
the pulse frequency from the atomic resonance in free space.
Averaged over a thermal ensemble, the Ramsey signal $P_\mathrm{Ramsey,inh}(t)$
is an integral over all allowed energies $E_a\leqslant h|U_0|$:
\begin{equation}
 \label{eq2}
P_\mathrm{Ramsey,inh}(t)=\int_0^{h|U_0|} p(E_a)P_0(t)dE_a.
\end{equation}
Given the measured $U_a$ and temperature $T_a$, we carry out
numerical evaluation of Eq.~(2) and obtain the values of $T_2^*$,
which is the $1/e$ decay time of the amplitude of Ramsey fringes.
%{\bf ??? I do not understand what it is that you have done below... rewrite so it is understandable... poor phrasing}
% 图3理论曲线是固定估计的T2'=300 ms, 通过计算不同势深下的T2*得到总的退相干时间\tao.
 At the magic light intensity ($U_a / U_M =1$) and a temperature of 17 $\mu$K, we obtain $T_2^*\approx$1.5 s. For different $U_a / U_M $ we thus have different $T_2^*$. Together with an estimated $T_2'\approx300$ ms~\cite{Yu2014}, and an independently measured value of $T_1\approx4$ s, the coherence time $\tau$ is deduced for each ratio of trap depths, and is plotted as a curve in Fig.\ref{fig:visi}.

 Notice that the predictions of described model deviate from the measurements when the trap depth is away from the magic point. This is likely caused by the neglected anharmonicity of the motion of the atoms in the Gaussian ODT at high temperatures.
In this experiment, the decay time of the Ramsey signal is dominated by the magnetic noise. It is worth noting that the homogeneous dephasing time due to relative intensity fluctuations ($0.15\%$) and heating rate ($2 \,\mu$K/s) are estimated to be
$300\, \mathrm{s}$ and $34\, \mathrm{s}$ respectively\cite{Kuhr2005,Yu2014}, thereby both of them can be neglected for magic trapping. Meanwhile, because of working magnetic bias field is relatively large, 3.115 G, compared to our previous work~\cite{Yu2014}, the sensitivity to the B-field noise is enhanced; this is presently the dominant  source of decoherence.

%%%%%%%%%%%%%%%%%%%%%%%%%%%
\begin{figure}[htbp]
\centering
\includegraphics[width=8.6cm]{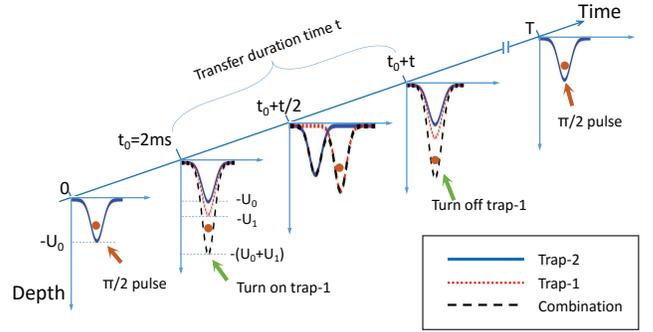}
\caption{(Color online) Schematic illustration of the transfer process of a mobile qubit. An atom in a superposition state (the qubit) is initially confined in the static ODT (trap-2). It is then overlapped with the mobile ODT (trap-1). The qubit is extracted out by the mobile ODT and becomes a mobile qubit. The mobile qubit travels for time interval $t$, and then it is returned to the static ODT. }
\label{fig:fig4}
\end{figure}
%%%%%%%%%%%%%%%%%%%%%%%%%%%%%

Finally, we study the coherence loss of a mobile qubit during a transfer process.
The key issue is to see whether the described magic trapping technique can mitigate the coherence loss of the mobile qubit.
The experimental time sequence is  illustrated in Fig.\ref{fig:fig4}.
The trap-1 (mobile ODT) and trap-2 (static ODT) serve as the ``moving head'' and the ``register'' respectively. The trap-2 is operated at the magic-intensity condition, i.e., trap depth of 0.17(2) mK and magnetic field of 3.115G.  In this trap, the measured temperature is about 8 $\mu$K, translating into $T_2^*\approx6.6$ s. Once the atom in the $|1\rangle$ state is confined in trap-2, a $\pi/$2 pulse is applied. At $1.9\, \mathrm{ms}$, the trap-1 is overlapped with trap-2, switched on, and ramped up to 0.2 mK within 0.1 ms. Then the trap-1 is moved away from trap-2 by linearly sweeping the AOD driving frequency. Since the moving trap-1 is deeper than trap-2, the atom follows trap-1~\cite{Yu2014} and is extracted out by the mobile ODT. The extracted atom becomes a mobile qubit. The mobile qubit travels for a duration time $t$. Then it is sent back to the static ODT, and the trap-1 is ramped down within 0.1 ms. The qubit returns to the original register site again. No measurable particle loss has been detected after the transfer process. To measure the coherence loss,  the second $\pi/2$ pulse is applied at  time $T$ to complete the Ramsey
interferometry sequence.

\begin{figure}[htbp]
\centering
\includegraphics[width=7.6cm]{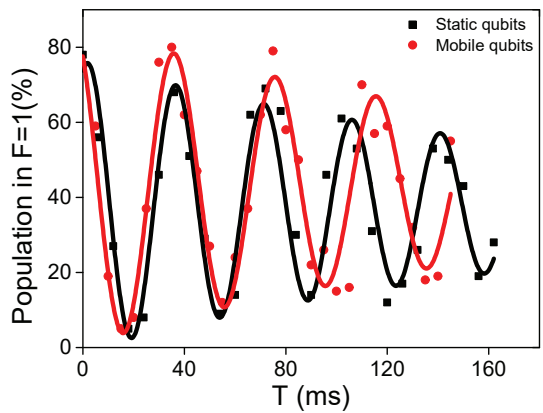}
\caption{(Color online) Measured Ramsey signals for single static qubits (black square) and single mobile qubits (red dots) at B=3.115 G. Every point is an average over 100 experimental runs. The solid curves are fits to the damped sinusoidal function. The fitted values of  coherence times $\tau$ of static qubits and mobile qubits are 206 $\pm$ 69 ms and 205 $\pm$ 74 ms respectively.}
\label{fig:Ramsey}
\end{figure}

The measured Ramsey signal as a function of time $T$ is shown in Fig.\ref{fig:Ramsey}, together with the Ramsey signal for static qubits. The fitted decay time of the Ramsey signal of single mobile qubits is the same as for the static qubits. At the beginning and the end of the transfer, the atoms are confined in an overlap of the two traps. The total trap depth is up to 0.37 mK and is far away form the magic operation condition. The dephasing time of the qubits trapped in this overlap trap is measured to be about 25 ms. But the actual trap overlap duration ($<$ 0.2 ms) is too short to cause significant dephasing. Besides,  for the measured temperature of 14 $\mu$K, the estimated dephasing time in the ``moving head'' trap is long, $T_2^*\approx3$ s . The entire transport takes only 2 ms and  the accompanying dephasing is negligible. After returning to trap-2, the temperature of the atoms is increased to 16 $\mu$K. Using Eq.~(2), the calculated $T_2^*$ in magic trap-2 drops to about 1.9 s because of the increase in the temperature. This causes mobile qubits to lose  $10\%$ of their coherence time, which is undetectable in the experiment, as verified by the data in Fig.~\ref{fig:Ramsey}. This is because with the magic trap method,
fluctuations of other sources like heating of atoms and pointing instabilities of the trap
laser beams have been greatly suppressed.
% {\bf which sources do you mean?}
The remaining dominant noise source is the magnetic noise which is not changed during the transfer process. The data in Fig.~\ref{fig:Ramsey} shows that mobile qubits do not experience additional coherence loss in the transfer process, and the magic ODTs is indeed robust for coherently transfer of mobile qubits.

%%%%%%%%%%%%%%%%%%%%%%%%%%%
%%%%%%%%%%%%%%%%%%%%%%%%%%%

In summary, we demonstrated a coherent transfer of a mobile qubit, a prerequisite step in realizing a large-scale neutral atom quantum information processing platform. This transfer was  crucially aided by magic trapping technique that mitigated the leading source of  decoherence, the differential light shift for two qubit states. To this end, we experimentally demonstrated the novel technique of magic intensity trapping. This technique relies on the importance of  the previously neglected ground state hyperpolarizability which makes the dependence of DLS on laser intensity parabolic; at the extrema of that dependence, the DLS is insensitive to spatial variations and fluctuations of the trapping laser intensity.  The measured coherence time is limited by the residual magnetic noise.
%The magic trapping technique can help in alleviating the detrimental heating effects in  manipulating single mobile qubits.
The coherence preservation of single mobile qubits has been demonstrated. Extending the operation to a large scale register is straightforward. Our results pave the way for constructing a scalable quantum-computing architectures with single atoms trapped in an array of ODTs. The quantum gate operation may also be improved by using the magic trapping technique~\cite{Xia2015}. Although this work has  focused on quantum information processing applications, the demostrated magic trapping technique is anticipated to benefit other studies with optically trapped atoms, e.g., controlled coherent collisions between $^{85}$Rb and $^{87}$Rb atoms~\cite{Xu2015}.

\begin{acknowledgments}
This work was supported  in part by the National Basic Research Program of China under Grant No.2012CB922101, the National Natural Science Foundation of China under Grant Nos.11104320 and 11104321,  funds from the Chinese Academy of Sciences and US National Science Foundation.
\end{acknowledgments}

%\bibliographystyle{apsrev4-1} % Tell bibtex which bibliography style to use
%\bibliography{referv4}

\end{document}